\documentclass[fleqn,usenatbib]{mnras}

\usepackage[T1]{fontenc}
\usepackage{ae,aecompl}

\usepackage{graphicx}	
\usepackage{amsmath}	
\usepackage{amssymb}	

\usepackage{newtxtext,newtxmath}

\title[GK Persei in 2010]{Timing and spectral analysis of GK Persei during the 2010 dwarf nova outburst}

\author[Pei et al.]{
Songpeng Pei,$^{1}$\thanks{E-mail: songpengpei@outlook.com}
Xiaowan Zhang,$^{1}$
Qiang Li,$^{2, 3}$
Ziwei Ou,$^{4}$
Taozhi Yang,$^{5}$
\newauthor
and Yongzhi Cai$^{6, 7, 8}$
\\
$^{1}$School of Physics and Electrical Engineering, Liupanshui Normal University, Liupanshui, Guizhou, 553004, China\\
$^{2}$Qiannan Normal University for Nationalities, Duyun 558000, China\\
$^{3}$Qiannan Key Laboratory of Radio Astronomy, Guizhou Province, Duyun 558000, China\\
$^{4}$Tsung-Dao Lee Institute, Shanghai Jiao Tong University, Shanghai 201210, China\\
$^{5}$Ministry of Education Key Laboratory for Nonequilibrium Synthesis and Modulation of Condensed Matter, School of Physics, Xi'an Jiaotong University,\\ 710049 Xi'an, China\\
$^{6}$Yunnan Observatories, Chinese Academy of Sciences, Kunming 650216, China\\
$^{7}$International Centre of Supernovae, Yunnan Key Laboratory, Kunming 650216, China\\
$^{8}$Key Laboratory for the Structure and Evolution of Celestial Objects, Chinese Academy of Sciences, Kunming 650216, China\\
}

\date{Accepted 00.00.2024. Received 00.00.2024; in original form 00.00.2024}

\pubyear{2015}

\begin{document}
\label{firstpage}
\pagerange{\pageref{firstpage}--\pageref{lastpage}}
\maketitle

\begin{abstract}
GK Persei, an old nova and intermediate polar (IP), exhibited a dwarf nova (DN) outburst in 2010. This outburst was extensively observed by the Neil Gehrels Swift Observatory, beginning 1.95 days after the eruption and continuing until 13.9 days before the maximum of the outburst in the optical. In this paper, we present timing and spectral analyses, comparing the results with those of other outbursts. We confirm the spin modulation in the 2 $-$ 10 keV X-ray range with a period of $P_{\rm WD} = 351.325(9)$ s. Additionally, we detected spin modulation in the 0.3 $-$ 2 keV band during the second half of the observations, a feature not seen in the 2015 and 2018 outbursts. This finding suggests that the soft X-ray emission in GK Per may originate partly near the magnetic poles and partly from a wind or circumstellar material.
\end{abstract}




\section{Introduction} \label{sec:intro}
GK Persei (A 0327+43, Nova Persei 1901) is a cataclysmic variable (CV) system consisting of a magnetized white dwarf (WD) \citep{1979MNRAS.187P..77K, 1983A&A...125..112B} and a K2-type subgiant star with mass of 0.25 -- 0.48 M$\odot$ \citep{1985MNRAS.212..917W, 2002MNRAS.329..597M, 2021MNRAS.507.5805A}. It is classified as an IP \citep{1985MNRAS.212..917W}. GK Persei underwent a classical nova explosion in 1901 \citep{1901ApJ....13..170P, 1901ApJ....13..173H, 1901MNRAS..61..337W} and was first observed to exhibit DN behavior in 1948, characterized by small-amplitude (1 $-$ 3 mag.) optical outbursts lasting 2 -- 3 months that may be due to the large physical size of its accretion disc \citep{2023MNRAS.523.3192M}. The recurrence time of these DN outbursts is irregular, typically ranging from 2 to 3 years, with most intervals just under 2.5 years \citep{1983A&AS...54..393S, 2002A&A...382..910S}. As an IP, GK Persei is a magnetic system, with a strong magnetic field around the WD estimated to be about 0.5 megagauss \citep{2018MNRAS.474.1564W}.

At a distance of 432$_{-7}^{+9}$ pc, derived from the parallax measurements in the {\sl Gaia} Early Data Release 3 (eDR3) \footnote{http://dc.g-vo.org/tableinfo/gedr3dist.main} \citep[][]{2021AJ....161..147B}, GK Persei is the second closest nova detected. It has been extensively studied both during outbursts and quiescence, and is home to the first classical nova remnant discovered in X-rays \citep{2005ApJ...627..933B}. Additionally, it hosts the largest known planetary nebula emitting X-rays in a WD binary system \citep{1989ApJ...344..805S}. The system’s orbital period is 1.9968 days \citep{1986ApJ...300..788C, 2002MNRAS.329..597M, 2021MNRAS.507.5805A}, one of the longest among CVs. The long DN outbursts in IPs are attributed to thermal-viscous instability in the accretion disk \citep{Hameury2017}. The orbital inclination angle, $i$, is estimated to be in the range of $63 - 73^{\circ}$ \citep{2018MNRAS.474.1564W, 2021MNRAS.507.5805A}. \citet{2016A&A...591A..35S} estimated the WD mass to be $M_{\rm WD} = 0.86~\pm~0.02~M_{\rm \odot}$, while \citet{2018MNRAS.474.1564W} proposed a WD mass of $0.87 \pm 0.08 M_{\odot}$ using the WD mass–radius relation. The primary star has a spin period of approximately 351 s \citep{1985MNRAS.212..917W} and a spin-up rate of $0.0003$ s yr$^{-1}$ \citep{2017MNRAS.469..476Z}. GK Persei has an estimated radius of $\sim$ 2 $\times$ 10$^{11}$ cm \citep{2009MNRAS.399.1167E}, with the suggestion that only a small fraction of disk is involved in a typical DN outburst, based on the disk's viscous decay timescale. \citet{2022A&A...659A.109K} proposed that the CO emission observed around GK Persei originates not from the circumstellar environment, but from the interstellar medium.

\citet{1991PASP..103.1149P, 1992MNRAS.254..647I} found a double-peaked modulation in the spin period during quiescence, while \citet{2004MNRAS.349..710H} reported a nearly sinusoidal modulation during outbursts. Additionally, both aperiodic and quasi-periodic oscillations (QPOs) with characteristic timescales of several kiloseconds have been detected in the X-ray and optical bands during outbursts \citep{1985MNRAS.212..917W, 1994ApJ...424L..57H, 1996MNRAS.283L..58M, 1999MNRAS.306..753M, 2002PASJ...54..987N, 2004MNRAS.349..710H, 2005A&A...439..287V}. QPOs with timescales of 380 s and 400 s were also detected in optical observations \citep{1981ApJS...45..517P, 1985A&A...149..470M}. \citet{1994ApJ...424L..57H, 2004MNRAS.349..710H} suggested that the frequently observed 5000 s QPO arises from bulges in the disk, caused by overflow as matter is accreted, moving with the local Keplerian timescale of approximately 5000 s at the inner disk radius.

During DN outbursts, the accretion rate increases compared to quiescence, causing the inner part of the accretion disk to move inward and pushing the magnetosphere toward the WD. The outer disk moves outward, and the magnetosphere radius, $R_{\rm m}$, increases when GK Persei is in quiescence \citep{2004MNRAS.349..710H, 2005A&A...439..287V, 2016A&A...591A..35S}. The maximum plasma temperatures during DN outbursts are lower than those in quiescence \citep{2009A&A...496..121B, 2017MNRAS.469..476Z, 2018MNRAS.474.1564W}, which is consistent with an expanding accretion disk \citep{2005A&A...435..191S, 2017MNRAS.469..476Z}.

The broadband X-ray spectrum in intermediate polars (IPs) is typically composed of a hard and a soft component. The hard X-ray component is highly absorbed \citep{1994PASJ...46L..81I}, and it can be explained by thermal bremsstrahlung radiation, where hot post-shock plasma cools as it interacts with the WD atmosphere. This post-shock plasma is generated by accreting material \citep{1973PThPh..49.1184A, 1999MNRAS.306..684C}. The soft X-ray emission is thought to arise from the reprocessing of the hard X-ray emission on the WD surface, originating in the accreting poles of the WD atmosphere, and is typically described by blackbody radiation \citep{1996A&A...310L..25B, 2004NuPhS.132..693D, 2004A&A...415.1009D, 2006A&A...454..287D, 2006A&A...449.1151D, 2009MNRAS.392..630L}. Some IPs, including GK Persei during its 2002, 2006, and 2015 outbursts, show the evidence for this blackbody component with temperatures ranging from 30 eV to 120 eV \citep{1994A&A...291..171H, 1996A&A...310L..25B, 2004NuPhS.132..693D, 2004A&A...415.1009D,2006A&A...449.1151D, 2006A&A...454..287D, 2007ApJ...663.1277E, 2008A&A...489.1243A, 2009MNRAS.392..630L, 2021ApJ...911...80M}. \citet{2003ApJ...586L..77M} showed that the soft X-ray spectra of GK Persei during outbursts are more consistent with a photoionization model rather than a cooling flow model.

The DN outbursts of GK Persei in 2010, 2015 and 2018 are similar, they are very different from the unusual 2006 outburst, which exhibited a smaller amplitude, a distinct duration, and multiple peaks, but they also display subtle differences. Thus, it is valuable to compare the 2010 outburst with the 2015 and 2018 outbursts. The optical light curves of the 2010, 2015, and 2018 outbursts each show a single peak at approximately 9.5, 9.8, and 10.1 magnitudes, respectively. In the 2010 outburst, the optical rise from its onset (MJD 55260.8) to its peak (MJD 55307.9) lasted 47.1 days. Swift XRT observations spanned 32.2 days, while the BAT data suggest the outburst ended on MJD 55348.0, resulting in a total estimated duration of 87.2 days. The 2015 outburst had a shorter optical rise, lasting 31.7 days from the beginning (MJD 57088.3) to the maximum (MJD 57120.0). Swift XRT observations covered 27.9 days, and BAT data indicate the outburst ended on MJD 57169, giving a total duration of 80.7 days. The 2018 outburst showed an optical rise lasting 33.0 days, from MJD 58352.0 to its peak at MJD 58385.0. Its total duration, as derived from the BAT data, was 67.9 days. Swift XRT observations of Epochs 1 and 2 began on MJD 58359.1 and ended on MJD 58380.0, covering a total of 20.9 days.

Due to its unique characteristics as a magnetic CV exhibiting both classical nova and DN activity, GK Persei has been studied extensively across various wavelengths. We analyzed optical, UV, and X-ray data from the 2010 DN outburst to investigate the evolution of its X-ray light curves and spectra. We identified a $\sim$ 351 s period in the 0.3 $-$ 2 keV light curve of the 2010 outburst, which differs significantly from the outbursts in 2015 and 2018 in which the period is detected only above 2 keV \citep{2017MNRAS.469..476Z, 2024MNRAS.529.1463P}. In this study, we perform timing and spectral analyses of GK Persei during the 2010 DN outburst, comparing it with the subsequent outbursts to explore the origin of the blackbody-like soft X-ray emission in this IP system.

\section{Observation and data reduction} \label{sec:observation}
The Neil Gehrels Swift Observatory \citep[hereafter, {\em Swift};][]{2004ApJ...611.1005G} began observing the 2010 outburst of GK Persei on March 7, 2010, 1.95 days after the eruption on March 5.8 \citep{2010ATel.2466....1E}, with observations lasting until April 8, 2010. The total observation duration was 32.19 days. Throughout the observation period, the Swift UV/Optical Telescope \citep[UVOT;][]{2005SSRv..120...95R} operated in image mode, providing the mean magnitude for each observation across one of the four UVOT filters (U, UVW1, UVM2, and UVW2). All UVOT data were processed using the FTOOLS package.

Swift X-ray Telescope \citep[XRT;][]{2005SSRv..120..165B} monitoring was carried out in both Photon Counting (PC) mode and Windowed Timing (WT) mode. The XRT light curves and spectra were generated using the XSELECT tool from the FTOOLS package in HEASOFT v. 6.30.1\footnote{https://heasarc.gsfc.nasa.gov/docs/software/lheasoft/download.html}. To minimize the effect of pile-up, the central region was excluded, and the XRTLCCORR command was applied for correction. Additionally, the XRT light curves were barycentrically corrected using the BARYCORR tool. Processed data from the Swift Burst Alert Telescope \citep[BAT;][]{2005SSRv..120..143B} were also incorporated, obtained from the Swift BAT transient monitor page \citep{2013ApJS..209...14K}. X-ray spectra were analyzed and fitted using XSPEC v. 12.12.1 \citep{1996ASPC..101...17A, 2003HEAD....7.2210D}.

A list of all the observations, including the observation date, exposure time, and mean count rate, is provided in Table~\ref{tab:obsgkper}.

\setcounter{table}{0}
\begin{table}
\begin{minipage}{80mm}
\caption[Log of all Swift XRT observations (in PC mode and WT mode)]{Log of all Swift XRT observations (in PC mode and WT mode) of GK Persei during its 2010 outburst.}
\label{tab:obsgkper}
\begin{tabular}{lccc}
\hline
\hline

      ObsID (PC mode)& Date$^{a}$ & Exp. (s)& Count rate  \\
\hline
00030842021 &55262.80  &1971  &1.87 $\pm$ 0.06  \\
00030842022 &55263.08  &1974  &1.51 $\pm$ 0.06  \\
00030842023 &55263.34  &1893  &2.05 $\pm$ 0.08  \\
00030842024 &55263.80  &1971  &2.12 $\pm$ 0.08  \\
00030842026 &55264.00  &24  &2.13 $\pm$ 0.56  \\
00030842027 &55264.01  &1716  &1.91 $\pm$ 0.08  \\
00030842028 &55264.41  &63  &2.11 $\pm$ 0.46  \\
00030842029 &55264.41  &1773  &2.01 $\pm$ 0.08  \\
00030842031 &55264.88  &924  &2.35 $\pm$ 0.10  \\
00031653002 &55265.21  &1752  &2.22 $\pm$ 0.08  \\
00031653004 &55265.55  &1704  &2.37 $\pm$ 0.08  \\
00031653006 &55265.95  &1338  &1.77 $\pm$ 0.07  \\
00031653008 &55266.15  &1758  &2.33 $\pm$ 0.08  \\
00031653009 &55266.49  &1581  &1.50 $\pm$ 0.07  \\
00031653010 &55266.96  &1161  &1.67 $\pm$ 0.08  \\
00031653011 &55267.49  &39  &1.59 $\pm$ 0.35  \\
00031653012 &55267.50  &1410  &1.87 $\pm$ 0.07  \\
00031653013 &55268.10  &114  &2.28 $\pm$ 0.06  \\
00031653014 &55268.10  &1239  &2.11 $\pm$ 0.08  \\
00031653015 &55268.23  &927  &2.28 $\pm$ 0.06  \\
00031653016 &55268.36  &996  &2.04 $\pm$ 0.09  \\
00031653017 &55269.03  &180  &1.41 $\pm$ 0.04  \\
00031653018 &55269.03  &1305  &1.20 $\pm$ 0.07  \\
00031653019 &55269.23  &1710  &1.41 $\pm$ 0.04  \\
00031653020 &55269.43  &1758  &1.56 $\pm$ 0.08  \\
00031653021 &55270.04  &66  &1.98 $\pm$ 0.06  \\
00031653022 &55270.04  &1224  &1.89 $\pm$ 0.08  \\
00031653023 &55270.37  &1524  &2.00 $\pm$ 0.07  \\
00031653024 &55273.04  &111  &1.10 $\pm$ 0.04  \\
00031653025 &55273.05  &1653  &0.99 $\pm$ 0.05  \\
00031653026 &55273.32  &1638  &1.21 $\pm$ 0.06  \\
00031653028 &55277.13  &108  &1.18 $\pm$ 0.19  \\
00031653029 &55277.13  &1776  &1.35 $\pm$ 0.06  \\
00031653030 &55277.33  &66  &1.10 $\pm$ 0.24  \\
00031653031 &55277.33  &1773  &1.33 $\pm$ 0.06  \\
00031653032 &55277.93  &42  &2.18 $\pm$ 0.67  \\
00031653033 &55277.94  &1764  &2.41 $\pm$ 0.09  \\

\hline

\hline

      ObsID (WT mode)& Date$^{a}$ & Exp. (s)& Count rate  \\
\hline
00030842021 &55262.79  &15  &2.41 $\pm$ 0.48  \\  
00030842022 &55263.08  &21  &1.64 $\pm$ 0.42  \\ 
00031653004 &55265.54  &30  &1.60 $\pm$ 0.26  \\ 
00031653013 &55268.09  &9  &2.90 $\pm$ 0.67  \\ 
00031653020 &55269.43  &21  &1.82 $\pm$ 0.32  \\ 
00031653021 &55270.04  &27  &1.95 $\pm$ 0.33  \\ 
00031653023 &55270.37  &18  &3.08 $\pm$ 0.48  \\ 
00031653026 &55273.32  &21  &1.24 $\pm$ 0.31  \\ 
00031653033 &55277.94  &18  &2.70 $\pm$ 0.44  \\ 
00031653034 &55280.01  &156  &1.74 $\pm$ 0.03 \\ 
00031653035 &55280.01  &2004  &1.70 $\pm$ 0.94  \\ 
00031653036 &55280.55  &114  &1.91 $\pm$ 0.03  \\ 
00031653037 &55280.55  &2007  &1.92 $\pm$ 0.98  \\ 
00031653038 &55280.88  &48  &1.67 $\pm$ 0.95  \\ 
00031653039 &55280.88  &2427  &1.32 $\pm$ 0.03\\ 
00031653040 &55284.03  &342  &2.34 $\pm$ 0.03  \\ 
00031653041 &55284.04  &2127  &2.36 $\pm$ 1.05  \\ 
00031653042 &55284.37  &342  &1.82 $\pm$ 0.03  \\ 
00031653043 &55284.37  &2010  &1.82 $\pm$ 0.03  \\ 
00031653044 &55284.89  &99  &1.16 $\pm$ 0.02  \\     
\hline
\multicolumn{4}{p{.9\textwidth}}{{\bf Notes: }$^a$ Modified Julian Date. The count rates were measured in the 0.3 -- 10.0 keV.}\\

\end{tabular}
\end{minipage} 
\end{table}

%
\setcounter{table}{0}
\begin{table}
\begin{minipage}{80mm}
\caption[Log of all Swift XRT observations (in PC mode and WT mode)]{Log of all Swift XRT observations (in PC mode and WT mode) of GK Persei during its 2010 outburst.}
\label{tab:obsgkper}
\begin{tabular}{lccc}
\hline
\hline

      ObsID (WT mode)& Date$^{a}$ & Exp. (s)& Count rate  \\
\hline
00031653045 &55284.89  &2322  &1.16 $\pm$ 0.02  \\ 
00031653046 &55287.10  &153  &1.46 $\pm$ 0.02  \\ 
00031653047 &55287.10  &5334  &1.46 $\pm$ 0.02  \\ 
00031653048 &55291.11  &15  &2.20 $\pm$ 1.08  \\ 
00031653049 &55291.12  &5331  &1.21 $\pm$ 0.02  \\ 
00031653050 &55294.39  &318  &1.11 $\pm$ 0.02  \\ 
00031653051 &55294.39  &1602  &1.13 $\pm$ 0.78  \\ 
00031653052 &55294.60  &1782  &1.25 $\pm$ 0.82  \\ 
00031653053 &55294.93  &1761  &1.01 $\pm$ 0.77  \\   
      
\hline
\multicolumn{4}{p{.9\textwidth}}{{\bf Notes: }$^a$ Modified Julian Date. The count rates were measured in the 0.3 -- 10.0 keV.}\\

\end{tabular}
\end{minipage} 
\end{table}

\section{Timing analysis}
Fig.\ref{fig:new hr} illustrates the comparison of the 2010 DN outburst of GK Persei across optical, UV, and X-ray energy bands. The top panel shows the optical light curve derived from data provided by the American Association of Variable Star Observers (AAVSO)\footnote{https://www.aavso.org/}. The onset (MJD 55260.8) and peak (MJD 55307.9) of the outburst in the optical band are indicated by vertical dotted lines across all panels as reference points. The second panel presents Swift UVOT observations through the U, UVW1, UVM2, and UVW2 filters. The third panel displays the Swift XRT light curve averaged per snapshot across the full energy range (0.3 -- 10 keV). Following the convention established by \citet{2017MNRAS.469..476Z, 2024MNRAS.529.1463P}, the 0.3 -- 2.0 keV and 2.0 -- 10 keV light curves are referred to as the soft and hard X-ray light curves, respectively. The fourth panel shows the soft and hard XRT light curves, while the fifth panel depicts the hardness ratio defined as (2 -- 10 keV)/(0.3 -- 2 keV). The sixth panel presents the Swift BAT light curve in the 15 $-$ 50 keV energy range. In this outburst, unlike in the 2018 and 2023 outbursts, the decay from maximum could not be followed in optical, UV and X-rays because the nova was close or behind the Sun.

To search for periodicities in the soft and hard X-ray light curves of GK Persei, we utilized Lomb-Scargle periodograms (LSPs) \citep{1982ApJ...263..835S} computed with the lomb package in R, developed by Thomas Ruf\footnote{https://cran.r-project.org/web/packages/lomb/index.html}. This package is designed for unevenly sampled time series, such as those with missing data. The false alarm probability level was set at 0.3\%. To validate the significance of the identified periodicities, we performed 100,000 random permutations of the data and computed the probability of random peaks exceeding the main peak in the original LSP. A Gaussian fit was applied to the primary peak in the LSP to estimate the 1$\sigma$ error of the derived periods.

\begin{figure*}
\begin{center}
\includegraphics[width=7.3cm]{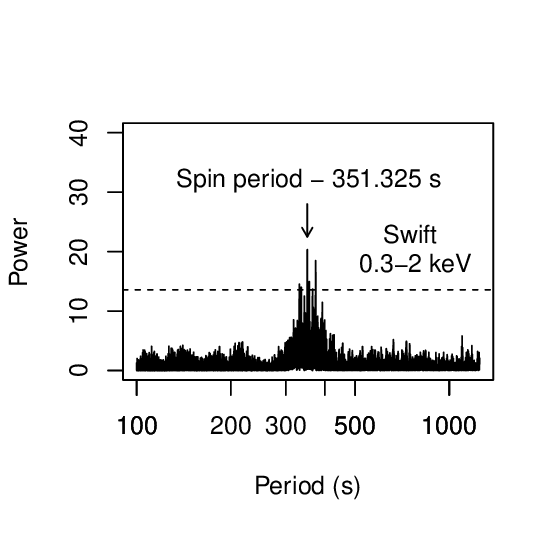}
\hspace{2.62em}
\includegraphics[width=7.3cm]{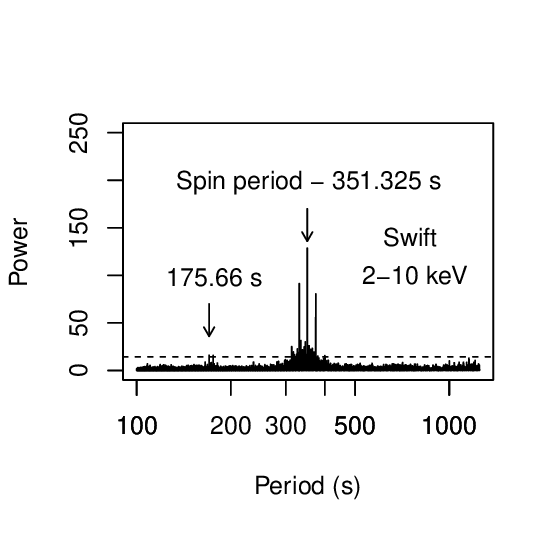}
\caption[The LSP plots]{Left: the Lomb-Scargle periodogram (LSP) derived from the 10-second binned light curve of Swift XRT data (the second half) in the 0.3 $-$ 2 keV energy range. Right: the LSP derived from the 10-second binned light curve of Swift XRT data in the 2 $-$ 10 keV energy range.

}
\label{fig:lsp}
\end{center}
\end{figure*}

\begin{figure*}
\begin{center}
\includegraphics[width=8.0cm]{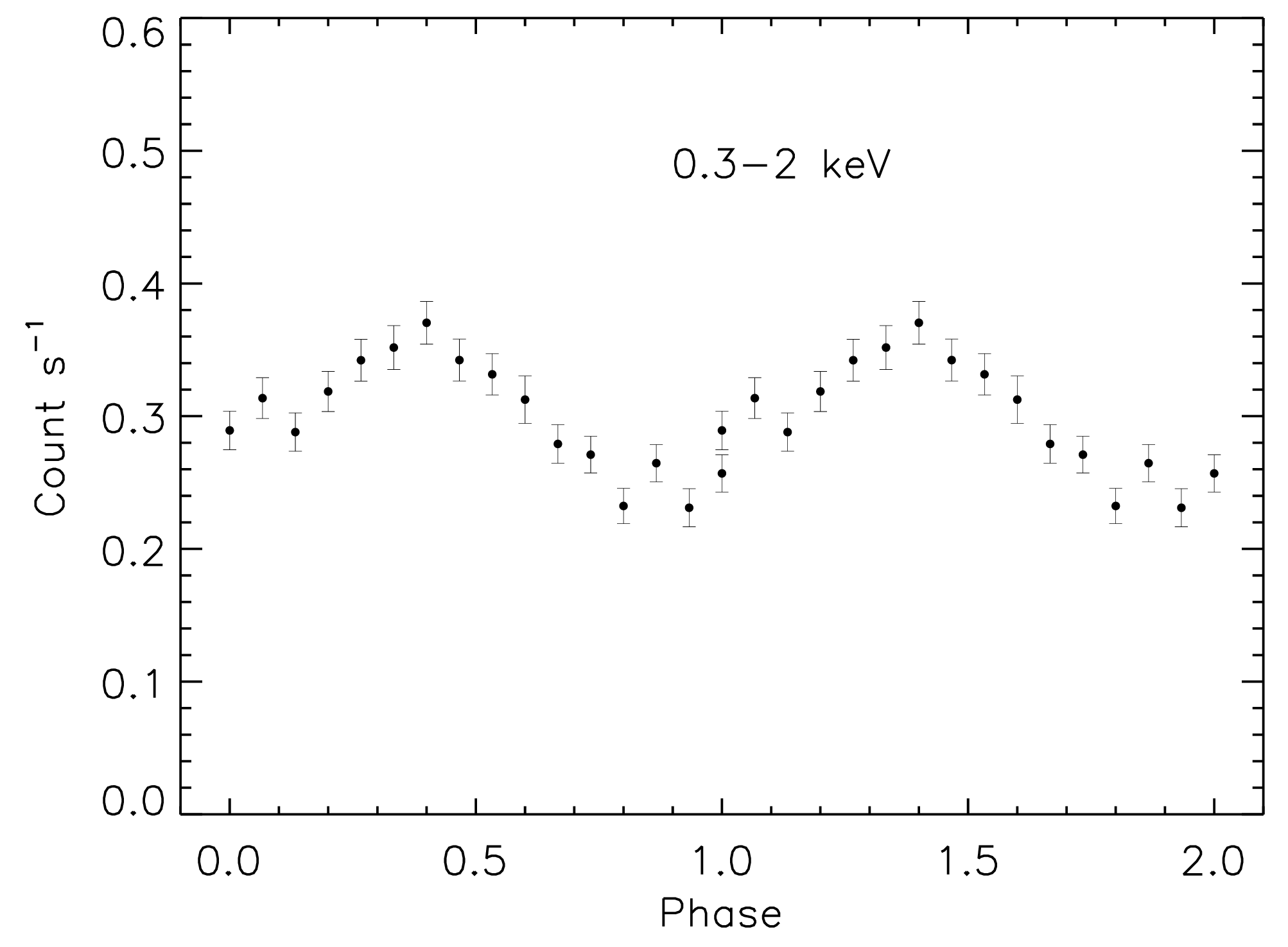}
\hspace{2.62em}
\includegraphics[width=8.0cm]{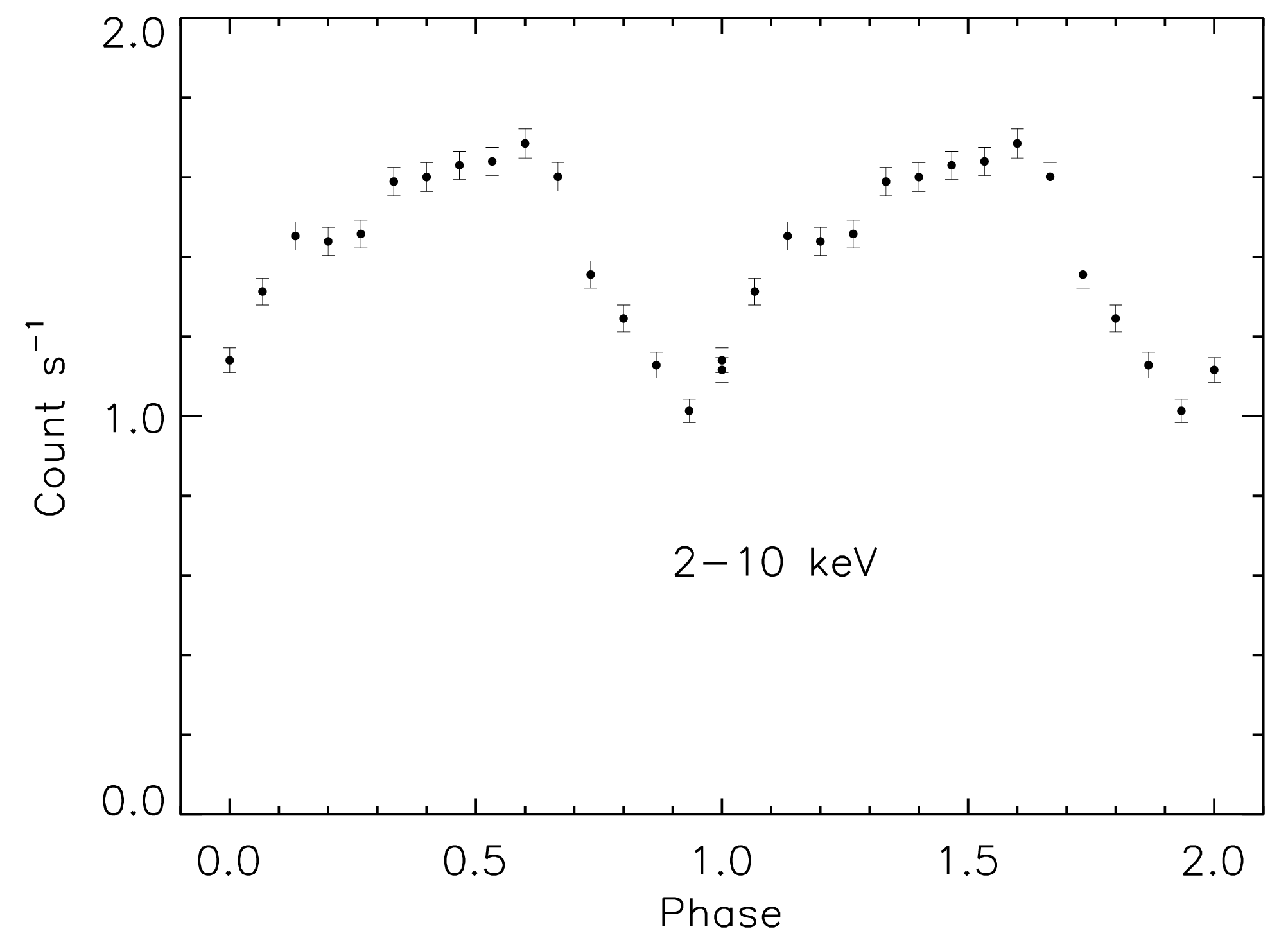}
\caption[The LSP plots]{Left: the 10-second binned light curve of the second half of the Swift XRT data in the 0.3 $-$ 2 keV energy range, folded using the detected period of 351.325 s. Right: the 10-second binned light curve of the Swift XRT data in the 2 $-$ 10 keV energy range, folded using the same detected period of 351.325 s.}
\label{fig:pfold}
\end{center}
\end{figure*}

Separate analyses of the 0.3 $-$ 2 keV and 2 $-$ 10 keV light curves were performed to enable comparisons with the results of \citet{2017MNRAS.469..476Z, 2024MNRAS.529.1463P}. During the 2015 and 2018 outbursts, the WD spin period was detectable only above 2 keV—a rare phenomenon among intermediate polars (IPs). We experimented with different bin sizes (10 s and 20 s) in the light curve analysis. While differences emerged for periods above 1000 s, only the peaks at $ 351.325 \pm 0.009$ s and $ 175.66 \pm 0.02$ s in the 2 $-$ 10 keV light curve and $ 351.325 \pm 0.011$ s in the 0.3 $-$ 2 keV light curve were consistent, stable, and independent of bin size. The $\sim$ 351.325 s period corresponds to the well-known WD spin period, and $\sim$ 175.66 s is its first harmonic. Notably, the $\sim$ 351.325 s period was observed only in the latter half of the observations (the second 16.10 days) in the 0.3 $-$ 2 keV band during the 2010 outburst. The corresponding LSPs for the soft and hard bands are shown in Fig.\ref{fig:lsp}, while Fig.\ref{fig:pfold} presents the spin-folded light curves for these bands. The folding process used a spin period of 351.325 s, with phase 0 defined as 2010-03-07 00:00:00.000 UTC (MJD 55262.0).

\begin{figure*} 
\includegraphics[width=15cm]{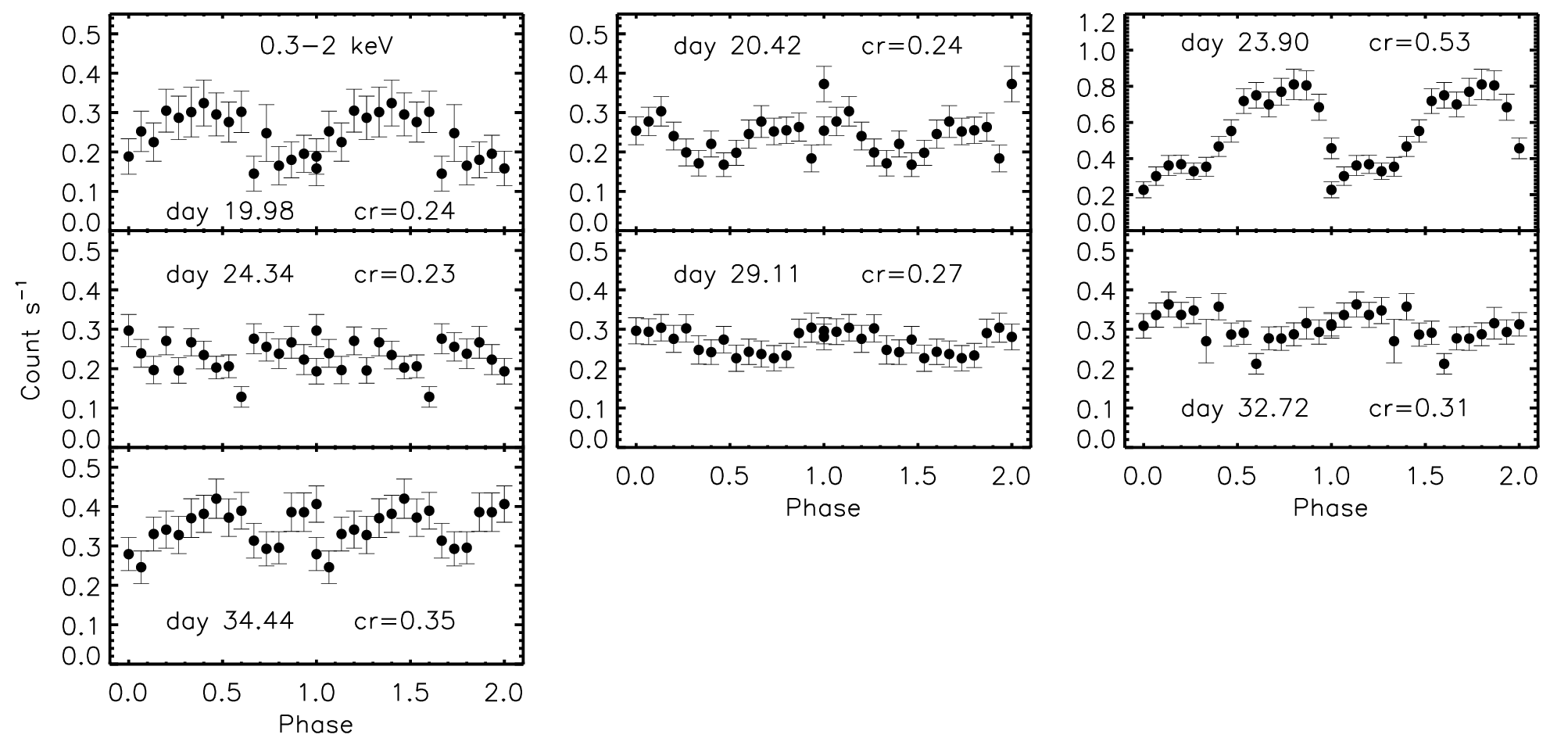}
\caption[Time evolution of spin-folded light curves]{Time evolution of the spin-folded light curves for the 2010 data (the second half observations) of GK Persei in the 0.3 $-$ 2 keV energy range. Each panel displays two spin cycles along the horizontal axis to enhance visibility. The mean observation date (measured in days since the eruption on March 5.8, 2010) and the mean count rate (cr) are indicated on each plot.}
\label{fig:pfold soft}
\end{figure*}

\begin{figure*} 
\includegraphics[width=15cm]{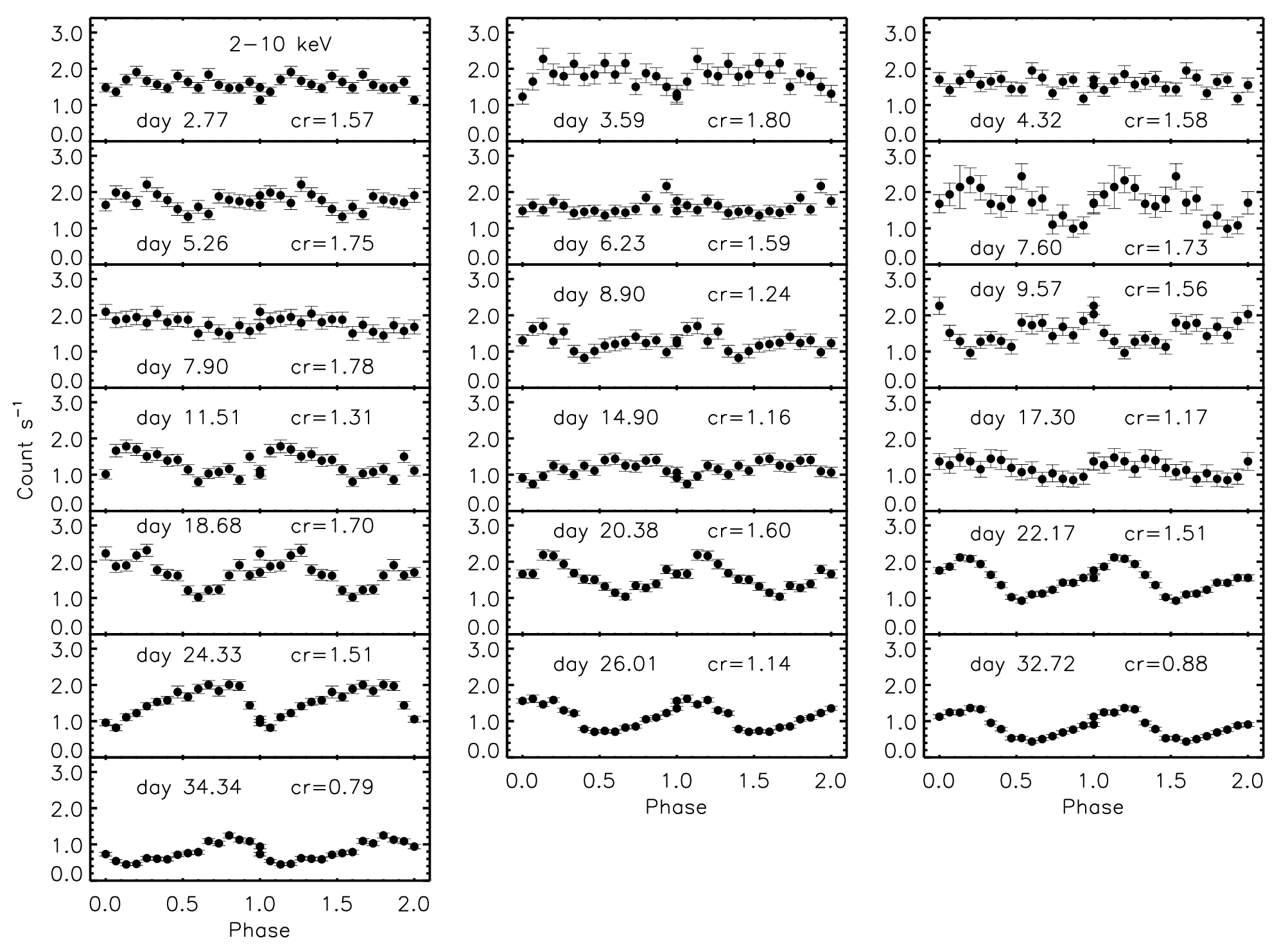}
\caption[Time evolution of spin-folded light curves]{Time evolution of the spin-folded light curves of GK Persei in the 2 $-$ 10 keV range from the 2010 data. Each panel displays two spin phases along the abscissa to enhance visibility. The mean observation date (measured in days since the eruption on March 5.8, 2010) and the mean count rate (cr) are indicated on each plot.}
\label{fig:pfold hard}
\end{figure*}

We examined the evolution of the spin pulse profile during the outburst. For the 0.3 $-$ 2 keV range, we grouped three individual observations (with exposure times lasting between 2274 s and 7251 s) from the second half of the light curve (where the period was detected) and folded them using the 351.325 s period (see Fig.\ref{fig:pfold soft}). For the 2 $-$ 10 keV range, the light curve for the entire observation period was similarly grouped (with exposure times lasting between 1572 s and 7809 s) and folded (see Fig.\ref{fig:pfold hard}). The pulse profile in the 2 $-$ 10 keV range becomes smoother over time, stabilizing approximately 20 days after the onset of the optical outburst. This behavior aligns with observations from the 2015 outburst \citet{2017MNRAS.469..476Z} but differs from those of the 2018 outburst \citep{2024MNRAS.529.1463P}. Notably, the spin period modulation in the 0.3 $-$ 2 keV range also becomes apparent around 20 days after the start of the optical outburst.

The period modulation amplitude is quantified as in \citet{2009MNRAS.399.1167E}, $(max-min)/(max+min)$. The period modulation amplitude increased from 26\% in the first half to 47\% in the second half of the 2 $-$ 10 keV light curve, lower than the 2018 outburst \citep{2024MNRAS.529.1463P} but with a smoother profile in 2010. For the second half of the 0.3 $-$ 2 keV light curve, the modulation amplitude was 34\%. We did not detect the $\sim$ 351.3 s period in the UV band due to the lack of event-mode data in the 2010 observations—all UV data were in image mode. This precluded confirmation of the tentative UV periodicity reported during the 2006 outburst \citep{2009MNRAS.399.1167E} but absent in the 2018 outburst \citep{2024MNRAS.529.1463P}.

The previously reported $\sim$ 5000 s period in X-rays (and sometimes in optical; \citealt{1996MNRAS.283L..58M, 2002PASJ...54..987N, 1985MNRAS.212..917W, 1994ApJ...424L..57H, 2005A&A...439..287V}) was not expected to be detected by Swift XRT due to its proximity to Swift’s orbital period (5754 s). Periodicities around this value, or its half (2877 s), were identified but varied with bin size and were deemed artifacts of the Swift orbit rather than intrinsic to the source.

\section{Spectral analysis}
The X-ray spectrum of GK Persei exhibits a high level of complexity. Timing analysis indicates the presence of at least two distinct sources of X-ray emission: one producing hard X-rays (2.0 -- 10 keV) and the other emitting in the soft X-ray range (0.3 -- 2.0 keV). The observed modulation suggests that the hard X-rays originate from a region obscured during the rotation period, likely near the WD poles, where accretion is funneled.

The spectrum above 2 keV is well-fitted by the cooling flow model \citep{1988ASIC..229...53M} (MKCFLOW model in XSPEC). Consistent with prior work, including \citet{2017MNRAS.469..476Z, 2024MNRAS.529.1463P}, the spectrum in the 0.3 $-$ 1.0 keV range is fitted with a blackbody (BB, or BBODY model in XSPEC). Since the elemental abundances are poorly constrained, we used the standard cooling flow model without varying abundances. The lower temperature in the MKCFLOW model was fixed at the minimum value allowed in XSPEC (0.0808 keV) to reduce free parameters. Other parameters are relatively insensitive to this choice.

For the 0.3 $-$ 2 keV range, the $\sim$ 351.3 s period is detected only in the second half of the observations. Consequently, spectra were extracted separately for the first and second halves, facilitating comparison between the 2015 and 2018 outbursts. Additionally, the first and second halves of the observations in 2010 correspond to the first and second two weeks of the Swift XRT observations in 2015, as well as to Epoch 1 and Epoch 2 of the Swift XRT observations in 2018, respectively. 

\begin{figure*}
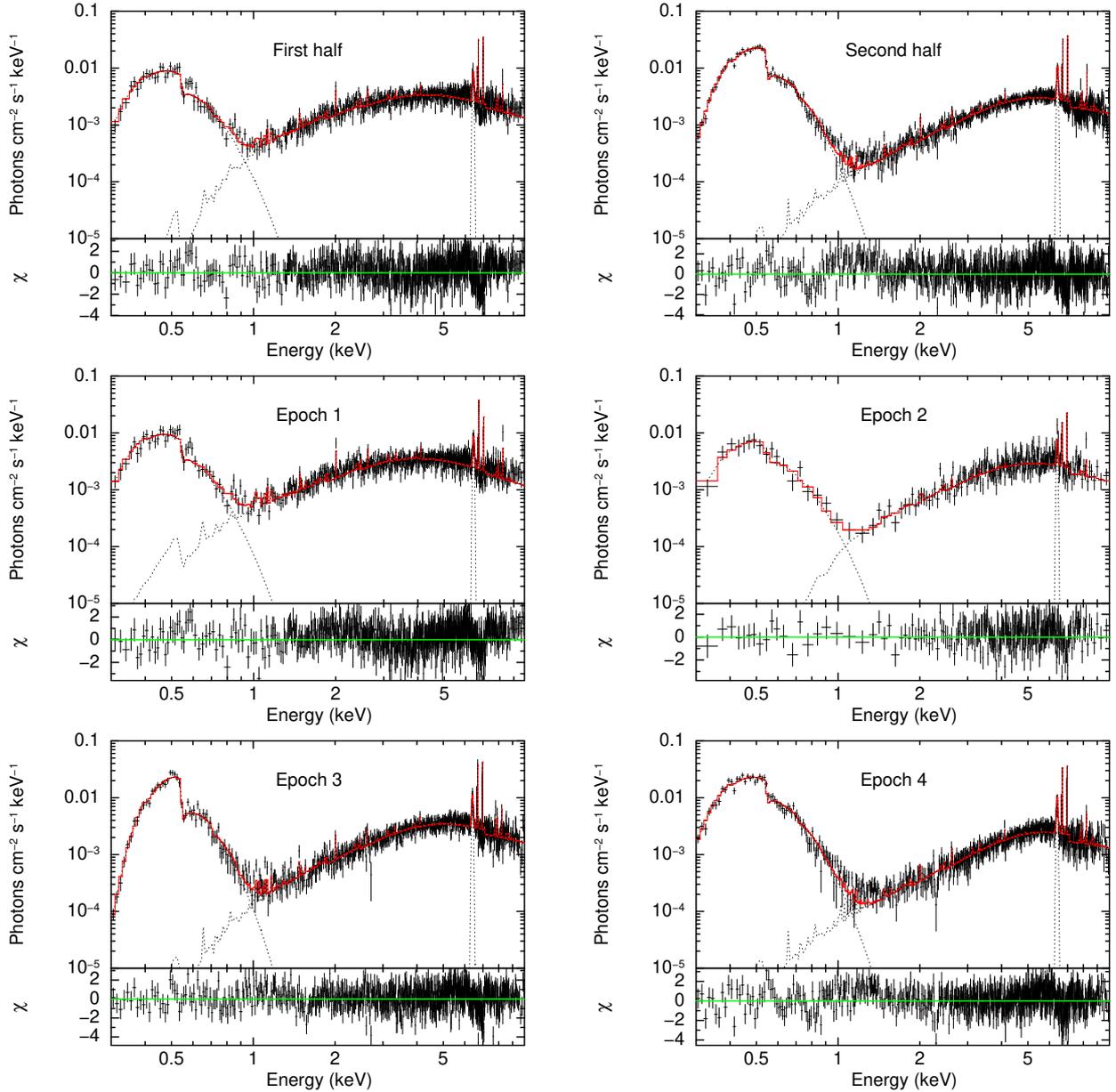

\begin{center}
\includegraphics[width=5.4cm,angle=270,clip]{final_first_half_01.eps}
\hspace{1.62em}
\includegraphics[width=5.4cm,angle=270,clip]{final_second_half_01.eps}
\includegraphics[width=5.4cm,angle=270,clip]{final_epoch_01.eps}
\hspace{1.62em}
\includegraphics[width=5.4cm,angle=270,clip]{final_epoch_02.eps}
\includegraphics[width=5.4cm,angle=270,clip]{final_epoch_03.eps}
\hspace{1.62em}
\includegraphics[width=5.4cm,angle=270,clip]{final_epoch_04.eps}

\caption[Unfolded spectra and the corresponding $\Delta \chi$ of the six epochs]{Unfolded spectra and the corresponding $\Delta \chi$ for the first and second halves of the observations, as well as the four epochs of GK Persei during the 2010 outburst, obtained using the XSPEC composite model \texttt{tbabs $\times$ (BB + pwab $\times$ (mkcflow + gaussian))}.}
\label{fig: Unfolded spectra}
\end{center}
\end{figure*}

Both spectra exhibit a prominent emission feature at approximately 6.4 keV, corresponding to the Fe K$\alpha$ fluorescent line, with no detectable energy shift (see Fig.~\ref{fig: Unfolded spectra}). A Gaussian component was added to model this line, fixing the central energy at 6.4 keV and the line width at 0.04 keV.

To account for interstellar medium (ISM) absorption, the Tuebingen-Boulder ISM absorption model (TBABS; \citealt{2000ApJ...542..914W}) in XSPEC was employed. Following \citet{2017MNRAS.469..476Z}, we augmented the column density N(H) model with a power-law distribution of neutral absorbers (PWAB in XSPEC) derived from the \texttt{wabs} code \citep{Done1998}. This addition was essential for obtaining a statistically acceptable fit to the MKCFLOW+Gaussian portion of the spectrum. Without it, the fit quality was inadequate.

Fig.\ref{fig: Unfolded spectra} shows the spectra and corresponding best fits, with parameter values summarized in Table~\ref{tab:model parameters}. Although efforts were made to incorporate an astrophysical plasma emission code (APEC; \citealt{2001ApJ...556L..91S}) to improve the high-resolution spectrum fit (as in \citealt{2017MNRAS.469..476Z}), no significant improvement was achieved in our case. Additionally, including a second MKCFLOW or BBODY component did not enhance the fit quality.

\begin{figure*}\begin{center}
\includegraphics[height=15.4cm]{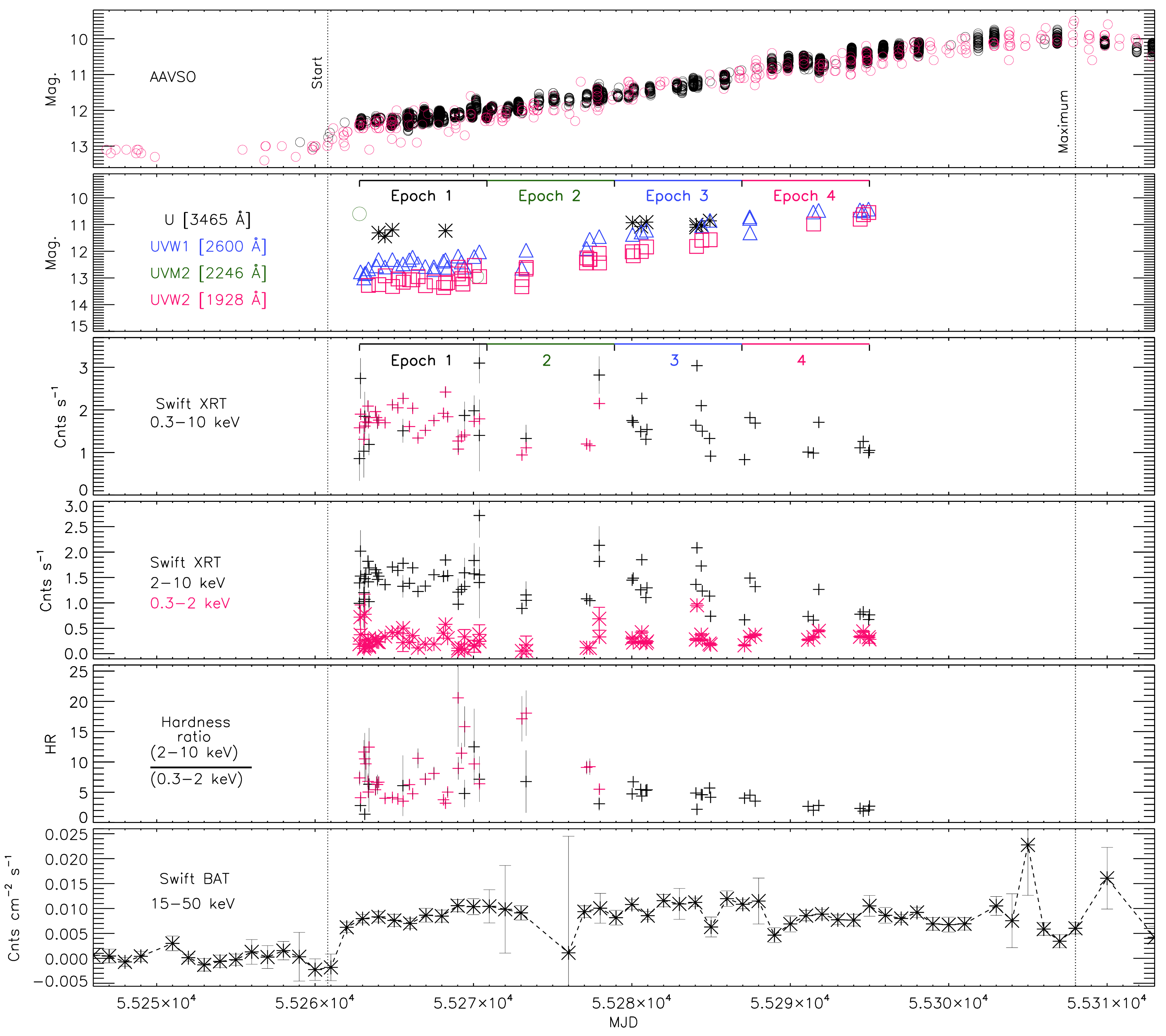}
\end{center}
\caption[Light curves in different filters]{From top to bottom: The AAVSO light curve of GK Persei is shown in the V band (black) and unfiltered (red). The two vertical dotted lines in all panels mark the onset (MJD 55260.8) and the peak (MJD 55307.9) of the 2010 outburst in the optical band. The subsequent panels display: the Swift UVOT light curves across various filters, the Swift XRT light curve in the 0.3 $-$ 10 keV energy band in PC mode (red) and WT mode (black), the Swift XRT light curves in the 2 $-$ 10 keV (black) and 0.3 $-$ 2 keV (red) bands, the X-ray hardness ratio [(2 $-$ 10 keV)/(0.3 $-$ 2 keV)], and the Swift BAT light curve. The horizontal lines with numbered labels in the second and third panels indicate four epochs defined for the analysis.}
\label{fig:new hr}
\end{figure*}

The fits did not constrain the maximum plasma temperature well. Since the total exposure time was sufficiently long, we divided the data into four time intervals to study the temporal evolution of the spectral shapes and explore better fits for the spectra. These intervals, labeled Epoch 1 to Epoch 4 (each spanning 8.05 days) in Fig.\ref{fig:new hr}, had exposure times of 34236, 8970, 13998, and 16296 seconds, respectively. The spectral fits showed a slight improvement with the reduced $\chi^2$ values decrease from 1.22 $-$ 1.26 to 1.08 $-$ 1.16 when applying the same model. The spectra and corresponding best-fit results for each epoch are also shown in Fig.\ref{fig: Unfolded spectra}, with the best-fit parameters listed in Table~\ref{tab:model parameters}.

An possible absorption feature at 0.76 keV (O VII) is evident in Epochs 1 and 4, while a more prominent absorption feature at $\sim$ 0.56 keV (O VII) appears in Epochs 3 and 4. These are indicative of warm absorber features \citep{2017PASP..129f2001M}. These two possible absorption feature was not detected in other outbursts of GK Persei. Similar complexities have been observed in other IPs. Warm absorber edges at 0.73 keV (O VII) and $\sim$ 0.9 keV (O VII) were reported in the Chandra HETG spectrum of V1223 Sagittarii \citep{2001ASPC..251...90M, 2021ApJ...919...90I}, marking the first unambiguous detection of a warm absorber in a CV. Similarly, the XMM-Newton RGS spectrum of V2731 Oph \citep{2008A&A...481..149D} revealed peculiar absorption components, including two high-density regions and a warm absorber with an O VII absorption edge at 0.74 keV. An absorption edge at 0.76 keV (O VII) has also been detected in IGR J08390-4833 \citep{2012A&A...542A..22B}.

An intriguing parameter of the MKCFLOW model is the mass accretion rate. In our fits, this parameter exhibited a slight increase from the rise to the maximum, that is different from the 2018 outburst, where it increased by a factor of two over the same phase. Despite dependencies on other fit parameters, this finding is significant. We are unable to precisely estimate the maximum plasma temperature as it exceeds the observational range of the Swift XRT.

%

%
\section{Discussion}
Fig.\ref{fig:new hr} illustrates that the UV and optical light curves share similar profiles. While the X-ray count rate above 2 keV shows some irregular variations, the hardness ratio initially increases before decreasing over time. This behavior contrasts with the 2015 and 2018 outbursts, where the hardness ratio declined steadily during the rising phase of the light curve, as the hard component flux increased \citep{2017MNRAS.469..476Z, 2024MNRAS.529.1463P}. In the 2010 outburst, the 0.3 $-$ 2 keV flux initially declined before rising, whereas in 2015 \citep{2017MNRAS.469..476Z} and 2018 \citep{2017MNRAS.469..476Z}, it increased steadily during the rise. 

Unlike the 2015 \citep{2017MNRAS.469..476Z} and 2018 \citep{2024MNRAS.529.1463P} outbursts, the $\sim$ 351 s periodicity in our data was detected not only in the 2 $-$ 10 keV energy range but also in the 0.3 $-$ 2 keV band during the second half of the observations. The spin modulation amplitude in the 0.3 $-$ 2 keV band is weaker than in the 2 $-$ 10 keV energy range. This suggests a variable origin of the emission in these two energy bands. The above phenomenon is puzzling, as in many IPs the modulation is thought to result from absorption, likely due to the "accretion curtain," where the effect of column density is more pronounced at lower energies \citep[e.g.][]{Rawat2022}. During the 2015 outburst, NuSTAR observations detected modulation even at higher energies (10 $-$ 40 keV). \citet{2017MNRAS.469..476Z} proposed that the greater height and wider extent of the accretion column in GK Per, compared to other intermediate polars (IPs) with shorter orbital separations, caused only moderate energy dependence due to absorption.

The temperature (T$_{\rm BB}$) of the blackbody component, derived from our fits, ranges between 69 and 84 eV and is consistent with values from other outbursts \citep{2005A&A...439..287V, 2007ApJ...663.1277E, 2009MNRAS.392..630L, 2009MNRAS.399.1167E, 2017MNRAS.469..476Z, 2024MNRAS.529.1463P}. However, this blackbody approximation is likely oversimplified, as the Swift XRT cannot resolve emission lines from ionized plasma. \citet{2024MNRAS.529.1463P} proposed that this component might not be a blackbody but is more likely ionized material emitting an emission-line spectrum, a conclusion with which we concur.

During the 2010 outburst, periodic modulation was detected in the 0.3 $-$ 2 keV band in the second half of the observations but was absent in this range during the 2015 and 2018 outbursts. \citet{2017MNRAS.469..476Z} analyzed Chandra HETG spectra from the 2018 outburst and found that between 0.7 $-$ 2.0 keV, the emission consisted only of emission lines. However, periodic modulation was present only in the Fe lines near 6.4 keV, not in other prominent lines. In 2010, the Fe line flux remained stable during Epochs 1 and 2 but increased by a factor of 3.2 from Epoch 2 to Epoch 3, while the total 2 $-$ 10 keV flux increased only slightly (1.1 times). This suggests that Fe lines and the remaining emission in this band originate from distinct materials. The soft X-ray flux (0.3 $-$ 2 keV) decreased from Epoch 1 to Epoch 2, likely due to increased absorption, and then doubled from Epoch 2 to Epoch 3, driven primarily by a 2.1 times increase in the blackbody component flux. The emitting area of the blackbody component increased initially before decreasing over time. Soft X-ray flux was higher in 2010 than in 2018, suggesting new emission in Epoch 3 compared to Epoch 2. This emission could originate near the poles or result from the reprocessing of hard X-rays. While \citet{2024MNRAS.529.1463P} suggested that most soft X-rays are diffuse plasma from a disk wind or similar mass-loss phenomenon, our results indicate that during Epochs 3 and 4, some soft X-ray flux likely originates from accreted material as well.

A potential absorption feature at $\sim$ 0.56 keV (O VII) in the spectra and spin periodic modulation in the 0.3 $-$ 2 keV band appear in Epochs 3 and 4 but not in Epochs 1 and 2. These findings support the hypothesis that absorption drives the observed modulation.

Throughout the outburst, the 2 $-$ 10 keV modulation amplitude in 2010 was larger than during quiescence, but less pronounced than in the 2015 event and comparable to 2018, though with slightly lower amplitude and smoother profiles. Spectral fits (Table~\ref{tab:model parameters}) estimate a mass accretion rate of 1.2 -- 1.8 $\times$ 10$^{-9}$ M$_\odot$ yr$^{-1}$ in 2010, compared to 0.6 -- 2.6 $\times$ 10$^{-8}$ M$_\odot$ yr$^{-1}$ during the 2015 rise \citep{2017MNRAS.469..476Z} and 1.1 -- 1.5 $\times$ 10$^{-9}$ M$_\odot$ yr$^{-1}$ during the 2018 rise \citep{2024MNRAS.529.1463P}. In the 2010 outburst, our fit did not contain the maximum plasma temperature well, and it is fixed at the 28 keV, therefore, we may have excessively estimate the value of the mass accretion rate. There is weak evdience that the mass accretion rate increased towards maximum, that is consistent with the result in 2018 \citep{2024MNRAS.529.1463P}. These results align with a positive correlation between pulsed fraction and mass accretion rate \citep{2024MNRAS.529.1463P}. We also note that the modulation amplitude increased toward maximum light in the 2010 outburst, that is also found in the 2018 outburst \citep{2024MNRAS.529.1463P}.

The hydrogen column density of the total absorber in the 2010 outburst is $\sim$ 0.18 -- 0.22 $\times$ 10$^{22}$ cm$^{-2}$, consistent with the interstellar absorption value of 2.17 $\times$ 10$^{21}$ cm$^{-2}$ \citep{2016A&A...594A.116H}. \citet{2012A&A...542A..22B} found that, for nine IPs, the total absorber's hydrogen column density was consistent with interstellar values, while the partial absorber reached up to 10$^{23}$ cm$^{-2}$ during outbursts. This agrees with the 2015 \citep{2017MNRAS.469..476Z} and 2018 \citep{2024MNRAS.529.1463P} outbursts of GK Per, as well as the 2010 outburst estimated by this work as shown in Table~\ref{tab:model parameters}.


\begin{table*}
\centering
\begin{minipage}{158mm}
\caption[The best fitting model parameters of the {\sl Swift} XRT data]{The best-fitting model parameters for the {\sl Swift} XRT data of GK Persei during the 2010 outburst, using the model \texttt{tbabs $\times$ (BB + pwab $\times$ (mkcflow + gaussian))}. The flux was determined using the CFLUX command in XSPEC. The quoted errors correspond to the 90\% confidence interval for a single parameter.}
\scalebox{0.99}{
\begin{tabular}{llcccccc}
\hline
Component & Parameter						&  \multicolumn{6}{c}{Value} 					\\
		  & 								&first half		  &second half	                          & epoch 1                       & epoch 2                         & epoch 3                                 & epoch 4             	\\
\hline
tbabs	  & nH$^{a}$ 						          &0.17 $_{-0.02}^{+0.03}$       &0.25 $_{-0.04}^{+0.03}$              &0.18 $_{-0.02}^{+0.02}$           &0.19 $_{-0.02}^{+0.01}$         &0.22 $_{-0.03}^{+0.02}$            &0.19 $_{-0.02}^{+0.02}$                             \\
\hline 
 		  & nH$_{\rm min}$$^b$  			          &0.25 $_{-0.12}^{+0.17}$       &0.19 $_{-0.09}^{+0.11}$              &0.17 $_{-0.06}^{+0.10}$            &53 $_{-12}^{+9}$                &0.16 $_{-0.07}^{+0.09}$            &0.13 $_{-0.05}^{+0.08}$                      \\    
pwab	  & nH$_{\rm max}$$^a$				                  &16 $_{-3}^{+5}$ 	         &31.6 $_{-10}^{+7}$                   &16 $_{-3}^{+4}$                   &34 $_{-9}^{+8}$                 &27 $_{-7}^{+6}$                    &41 $_{-8}^{+6}$    	\\    
		  & $\beta$						  &0.2 $_{-0.05}^{+0.06}$	 &0.39 $_{-0.12}^{+0.09}$              &0.21 $_{-0.05}^{+0.04}$           &0.31 $_{-0.08}^{+0.07}$         &0.39 $_{-0.10}^{+0.09}$             &0.47 $_{-0.14}^{+0.12}$    	       \\
\hline   
mkcflow   & T$_{\rm high}$ (keV)			                  &40 $_{-13}^{+11}$	         &48  $_{-15}^{+16}$                   &28 $_{-7}^{+6}$                   &28$^c$                          &28$^c$                                 &28$^c$                 \\
		  & $\dot{m}$$^d$					  &9 $_{-3}^{+2}$               &10  $_{-3}^{+3}$                     &12 $_{-3}^{+2}$                   &16 $_{-2}^{+2}$                 &18 $_{-4}^{+3}$                    &17 $_{-3}^{+4}$                   \\ 
\hline
		  & E (keV)$^c$						  & 6.4     		         & 6.4                                 & 6.4                              & 6.4                            & 6.4                               & 6.4                    \\
Gaussian  & $\sigma$ (keV)$^c$ 				                  & 0.04		         & 0.04                                & 0.04                             & 0.04                           & 0.04                              & 0.04                       \\
\hline  
BB		  & T (eV) 						  &78 $_{-13}^{+12}$  	         &69   $_{-8}^{+9}$                    &71 $_{-6}^{+7}$                   &84 $_{-8}^{+6}$                 &73 $_{-8}^{+8}$                    &76 $_{-9}^{+8}$                        \\
\hline    
Flux$_{\rm 0.3-2\ keV}$$^e$& abs.    		                          &4.2 $_{-0.7}^{+0.6}$	         &4.8  $_{-0.8}^{+0.8}$                &4.7 $_{-0.6}^{+0.4}$              &2.4 $_{-0.3}^{+0.4}$            &4.7 $_{-0.5}^{+0.6}$               &5.1 $_{-0.6}^{+0.6}$                  \\
						  & unabs. 		  &125  $_{-13}^{+17}$ 	         &205  $_{-22}^{+23}$                  &151 $_{-16}^{+13}$                &186 $_{-12}^{+11}$              &218 $_{-14}^{+14}$                 &206 $_{-16}^{+14}$                        \\
Flux$_{\rm 2-10\ keV}$$^e$ & abs.    		                          &190  $_{-18}^{+16}$	         &189  $_{-17}^{+15}$                  &188 $_{-13}^{+12}$                &178 $_{-14}^{+11}$              &205 $_{-15}^{+15}$                 &155 $_{-18}^{+12}$                     \\
						  & unabs. 		  &239 $_{-21}^{+23}$	         &304  $_{-28}^{+28}$                  &246 $_{-18}^{+15}$                &312 $_{-23}^{+24}$              &326 $_{-26}^{+24}$                 &305 $_{-28}^{+24}$                    \\						  
Flux$_{\rm mkcflow}$$^e$& abs.    		                          &186 $_{-35}^{+34}$	         &182  $_{-32}^{+37}$                  &184 $_{-26}^{+27}$                &173 $_{-24}^{+27}$              &198 $_{-26}^{+26}$                 &147 $_{-22}^{+23}$                  \\
						  & unabs. 		  &348 $_{-48}^{+45}$	         &434  $_{-42}^{+41}$                  &377 $_{-35}^{+36}$                &485 $_{-43}^{+37}$              &498 $_{-46}^{+45}$                 &463 $_{-49}^{+46}$                       \\
Flux$_{\rm Gaussian}$$^e$& abs.    		                          &6.5 $_{-1.9}^{+1.5}$	         &7.9  $_{-2.2}^{+1.8}$                &6.5 $_{-0.9}^{+1.2}$              &5.2 $_{-1.1}^{+1.1}$            &8.2 $_{-1.9}^{+1.6}$               &8.1 $_{-1.9}^{+1.6}$                  \\
						  & unabs. 		  &6.9 $_{-2.1}^{+1.8}$	         &8.5  $_{-2.4}^{+2.1}$                &7.1 $_{-1.7}^{+1.4}$              &5.9 $_{-1.2}^{+1.3}$            &9.1 $_{-1.8}^{+1.7}$               &9.6 $_{-1.9}^{+1.8}$                       \\
Flux$_{\rm BB}$$^e$& abs.    		                                  &1.8 $_{-1.1}^{+1.1}$	         &4.0   $_{-2.7}^{+2.4}$                 &1.8 $_{-1.0}^{+0.9}$                &1.6 $_{-0.7}^{+0.8}$            &3.4 $_{-1.6}^{+1.5}$               &4.6 $_{-2.5}^{+2.3}$                  \\
						  & unabs. 		  &12 $_{-2}^{+2}$	         &66   $_{-5}^{+5}$                    &15 $_{-2}^{+2}$                   &11 $_{-2}^{+4}$                 &37 $_{-3}^{+3}$                    &38 $_{-3}^{+3}$                        \\
\multicolumn{1}{l}{L$_{\rm BB}$ ($\times10^{33}$erg s$^{-1}$)}   &        &0.61 $_{-0.15}^{+0.16}$       &4.31  $_{-0.67}^{+0.49}$             &0.94 $_{-0.09}^{+0.10}$            &0.49 $_{-0.09}^{+0.09}$         &2.17 $_{-0.38}^{+0.32}$            &2.02 $_{-0.33}^{+0.31}$                     \\
\multicolumn{1}{l}{R$_{\rm BB}$$^f$ ($\times10^5$cm)}	    &             &11.2 $_{-2.2}^{+2.3}$         &38.5  $_{-4.3}^{+3.1}$               &16.6 $_{-2.5}^{+1.9}$             &9.0 $_{-1.5}^{+1.6}$              &28.6 $_{-2.5}^{+2.2}$              &21.7 $_{-2.1}^{+1.8}$                   \\			
\multicolumn{1}{l}{L$_{\rm 2-10\ keV}$ ($\times10^{33}$erg s$^{-1}$)} &   &5.3 $_{-0.3}^{+0.3}$	         &6.8 $_{-0.5}^{+0.4}$	               &5.5 $_{-0.2}^{+0.2}$              &7.0 $_{-0.3}^{+0.3}$              &7.3 $_{-0.3}^{+0.3}$               &6.8 $_{-0.2}^{+0.2}$ 		\\
\multicolumn{1}{l}{$\chi^2$}				   &              &1.22		                 &1.26                                 & 1.13	                          & 1.08                           & 1.15                              & 1.16                                           \\
\hline
\multicolumn{8}{p{.9\textwidth}}{{\bf Notes}: $^a$$\times$10$^{22}$ cm$^{-2}$. $^b$$\times$10$^{20}$ cm$^{-2}$. $^c$Frozen parameter. $^d$Mass accretion rate $\times$10$^{-10}$ M$_\odot$ yr$^{-1}$. $^e$$\times10^{-12}$ erg cm$^{-2}$ s$^{-1}$.}\\
\multicolumn{8}{p{.9\textwidth}}{$^f$Radius of the emitting region. The distance is assumed to be 432 pc.}\\
\hline 
\end{tabular}
}
\label{tab:model parameters}
\end{minipage}
\end{table*}

\section{Conclusions}
GK Persei is a classical nova that provides a unique opportunity to study a wide range of phenomena due to its relatively close distance. It is an unusual system, often referred to as a "Rosetta stone" for WD binaries. Studying GK Persei allows us to gather critical information about fundamental physical parameters, aiding the development of physical models. Through X-ray and UV observations of the 2010 outburst, we made significant progress in understanding the IP nature of this system, particularly its rare, DN-like outburst behavior in a magnetic CV that also undergoes classical nova events. Below are the key conclusions drawn from this eruption.

The WD spin period of $351.325 \pm 0.009$ s was detected in the 2 -- 10 keV range during the 2010 outburst, though the modulation was less prominent compared to the 2015 outburst. We confirm the correlation between the pulsed fraction and the mass accretion rate, as reported by \citet{2024MNRAS.529.1463P}. The spin modulation in the hard X-ray flux (above 2 keV) indicates that the hard X-rays originate from the accretion columns funneling onto the poles.

We also detected spin modulation in the softer energy band (0.3 $-$ 2 keV) light curve of the second half observations, but with a weaker amplitude than in the 2 -- 10 keV range. This suggests a larger shock height in GK Persei than in other IPs, where the hard X-ray flux is either smaller or not as modulated by the WD spin. The presence or absence of pulsations in the soft X-ray band indicates that the soft, blackbody-like component may partially originates from a small hot spot near the poles and partially originates from diffuse thermal plasma, possibly from a disk wind or a similar mass loss phenomenon. The soft, blackbody-like component may mostly originate from diffuse thermal plasma when the spin modulation in the soft X-ray band is absent. Additionally, we find evidence of warm absorber features in the spectra of GK Persei during the 2010 outburst and suggest that the spin modulation detected in the softer energy band may be influenced by absorption.

Although we could not precisely estimate the maximum plasma temperature, our spectral fits support the conclusion that the maximum plasma temperature may have varied irregularly during the outburst proposed by \citet{2024MNRAS.529.1463P}.

Finally, we found that the mass accretion rate varies significantly in the different DN-like outbursts of GK Persei. The values we obtained from spectral fits for the 2010 outburst, as well as the 2018 outburst \citep{2024MNRAS.529.1463P}, were about an order of magnitude lower than those derived for the 2015 outburst by \citet{2017MNRAS.469..476Z}.

\section*{Data Availability}
The data analyzed in this article are all available in the HEASARC archive of NASA at the following URL: \url{https://heasarc.gsfc.nasa.gov/db-perl/W3Browse/w3browse.pl}

\section*{Acknowledgements}
We extend our sincere gratitude to the anonymous referee for her or his insightful and constructive comments, which helped us to improve the scientific content of this article. We extend our gratitude to observers worldwide for their valuable contributions to the AAVSO database. Additionally, this research utilized data provided by the UK Swift Science Data Centre at the University of Leicester. This work is supported by the High-level Talents Research Start-up Fund Project of Liupanshui Normal University (Grant No. LPSSYKYJJ202208), Science Research Project of University (Youth Project) of the Department of Education of Guizhou Province (Grant No. QJJ[2022]348, QJJ[2022]346), the Science and Technology Foundation of Guizhou Province (Grant No. QKHJC-ZK[2023]442), the Discipline-Team of Liupanshui Normal University (Grant No. LPSSY2023XKTD11), the Liupanshui Science and Technology Development Project (Grant No. 52020-2024-PT-01), the Research Foundation of Qiannan Normal University for Nationalities (Grant No. QNSY2019RC02), the National Natural Science Foundation of China (Grant No. 12393853, 12003020, 12303054), the Shaanxi Fundamental Science Research Project for Mathematics and Physics (Grant No. 23JSY015), the Yunnan Fundamental Research Projects (Grant No. 202401AU070063), and the International Centre of Supernovae, Yunnan Key Laboratory (Grant No. 202302AN360001).




\bibliographystyle{mnras}
\bibliography{GKPer} 




\bsp	
\label{lastpage}
\end{document}